# Geometric phase associated with Poincaré beams due to unfolding of fractional optical vortex beams

**Satyajit Maji, Aswini K. Pattanayak and Maruthi M. Brundavanam**
Department of Physics, Indian Institute of Technology Kharagpur, Kharagpur 721302, West Bengal, India.

Email: bmmanoj@phy.iitkgp.ac.in

**Abstract**: Fractional optical vortex beam is generated by the diffraction of a Gaussian beam using computer generated hologram embedded with mixed screw-edge dislocation. Unfolding of the generated fractional vortex beam into eigen-polarization components with orthogonal polarization results in the conversion of scalar phase singularity to vector polarization singularities in the beam cross-section. The evolution of the singularities of the ellipse field namely C-points (points of undefined major axis) and L-lines (lines of undefined handedness) in the state of polarization distribution on a transverse plane quantifies the transformation. The effect of the phase morphology dictated by the fractional order of the dislocation, transverse spatial separation and longitudinal relative phase of the two eigen-polarization components on determining the complex transverse polarization structure is investigated. The nature of the generated Poincaré beam is also indicated by projecting the states of polarization on to the Poincaré sphere. With increasing order of dislocation from 0.0 to 1.0 in fractional steps and with increasing relative phase, the partial Poincaré beam is transformed to a full Poincaré beam. The transformation of the local structure around the C-points is measured through the geometric phase due to the Poincaré sphere contour around the C-points for different dynamic phase difference of the unfolded FOV beams generated from different fractional order of dislocation. This study can be useful for different geometric phase based application of optical vortex beams.

## I. INTRODUCTION:

A singularity, the smallest structure of an optical field, is of utmost important in nanoscale applications e.g. on-chip photonic components and in general light-matter interaction at the nanoscale [1-4]. The singularity in the optical field can be a scalar singularity where phase is undefined and intensity is zero or a vector or ellipse field singularity where some parameters quantifying the polarization state is undefined [3, 5]. Studies have been carried out to generate and classify both the scalar and vector singularities to enhance their utility and understanding [1-22]. Scalar optical vortex (OV) beams with nested phase singularities that carry orbital angular momentum (OAM) have found applications in optical trapping and manipulation of micro-particles [9], Super-resolution imaging (STED) [10], robust, high capacity, secure optical communication [11, 12], hybrid entanglement [13] among many others. Their studies have enabled detailed understanding of optical current flow [14] and optical angular momentum [15]. The beams with polarization singularities are used in engineering the point spread function for imaging [16] and to enhance the sensitivity in polarimetry methods [17], etc., to name a few [18]. Continuing investigation of optical fields embedded with both type of singularities has initiated new fields of study in optical physics like spin-orbit coupling [19], spin and orbital Hall effect [20, 21], chiral light matter interactions [22] to give some examples from a long list.

The strength of the singularity is characterized by signed numbers called topological charge (TC) for scalar singularities and topological index (TI), Poincaré-Hopf index for ellipse and vector singularities [3]. Scalar OV beams generated from fractional helicoidal phase steps have exclusive topological features which have been theoretically and experimentally investigated [23-25]. The fractional OV (FOV) beams have intricate optical current flow across the beam and both extrinsic and intrinsic OAM [25-28]. The FOV beams have an overall and near-core anisotropic field distribution [25, 29], and a broad OAM spectrum which is useful in high-dimensional entanglement [30], digital spiral imaging [31] and novel trapping applications [32].

On the other hand, a spatially inhomogeneously polarized vortex beam contain singularities of ellipse field in a transverse plane, where the major axis and handedness are undefined and are known as C-points and L-lines respectively [3]. In the parametric space of the Poincaré sphere, if the projection of the state of polarization (SoP) across an optical beam spans the entire surface of the Poincaré sphere, the beams are called Poincaré beams [33]. The mutual relation and conversion of the scalar and polarization singularities [34], the classical non-separable nature where two different degrees of freedom like spatial mode and polarization are entangled [35] and quantum properties of such vector vortex beams [36] are some recent interests. In this paper we consider the transformation of scalar singularities to polarization singularities using FOV beams.

For a vector optical field, the inhomogeneous polarization distribution can be associated with two phases, one that describes the instantaneous electric field direction with respect to the major axis of the ellipse, tip of which traces the ellipse. And another phase that describes the orientation of the major axis of the ellipse field. The former is the integrable dynamic phase (DP) which is quantized when integrated along a contour. The latter is non-integrable and unbounded geometric phase (GP) that is not quantized [37]. As the origin of this GP lies in the manipulation of the SoP of the light, it is also called the Pancharatnam-Berry phase. The GP is most important in the context of Poincaré beams and provides a mean of probing the field near the singular points of the ellipse field. Also because of more robustness of GP than DP, it is very useful in the quantum and other applications [38] in science and is the underlying reason for the spin-orbit interactions [39] and effects arising from interaction of polarized light with nanostructures [40] and meta-materials [41].

In the current investigation, we have exploited the anisotropy of scalar FOV beams to generate and characterize the Poincaré beams. The scalar FOV beams are first generated using fractional spiral phase encoded mixed screw-edge dislocation (MSED) holograms or fork-cut holograms. The generated scalar FOV beams are then passed through a uniaxial birefringent ($YVO_4$) crystal to convert them to polarization singularities via unfolding due to birefringence [34, 42-44]. The SoP distribution in the transverse plane is obtained from the measurement of spatially resolved Stokes parameters [34]. From the measured Stokes parameters, three scalar Stokes fields are computed and the polarization singularities are described in terms of the singularities of the Stokes fields [42, 43].

The scalar to polarization singularity transformation is demonstrated by the evolution of the C-points and L-lines across the Poincaré beam. The effect of the determining parameters like phase morphology described by the TC [25, 45], beam separation and relative phases of the eigen-polarization components are exclusively studied and discussed. The GP associated with the spatial polarization around the C-point singularities due to unfolding of different fractional vortex beams are measured. The variation of the GP due to orientation of the optic axis of the crystal is demonstrated by controlled rotation of the birefringent crystal. The fine-tuning and measurement of the dynamic relative phase is achieved by employing the idea originally employed for the weak measurement of birefringence induced very small spatial walk-off [46].

## II. THEORY:

**A. Fractional OV beams:**
The optical field of an integer TC (m) vortex generated by the diffraction of a Gaussian beam from a computer generated Hologram (CGH) with point screw dislocation can be written as [25, 47],

$$u_m(\rho,\phi,z) = \sqrt{\frac{\pi}{2}} e^{im\phi}(-i)^{|m|+1} \exp(\frac{ik}{2z}\rho^2)\frac{z_R}{z-iz_R}\sqrt{A}\exp(-A)\left[I_{\frac{|m|-1}{2}}(A) - I_{\frac{|m|+1}{2}}(A)\right] \quad (1)$$

where $I_\nu$ denotes the modified Bessel function of first kind of order $\nu$, $z_R = \frac{\pi \omega_0^2}{\lambda}$ is the Rayleigh range and $A = \left(\frac{\rho}{z}\right)^2 \frac{kz_R}{4(1-i\frac{z_R}{z})}$.

CGHs embedded with fractional spiral phase giving rise to FOV beams in the far-field. The resulting fractional order $(\alpha)$ OV field expressed as a superposition of integer charge vortex fields is given as [23].

$$u_\alpha(\rho,\phi,z) = \frac{\exp[i(z+\pi\alpha)]\sin(\pi\alpha)}{\pi} \sum_{m=-\infty}^{\infty} \frac{u_m(\rho,\phi,z)}{(\alpha-m)} \quad (2)$$

**B. Spatial separation and relative phase due to unfolding inside a birefringent crystal:**
In a uniaxial birefringent crystal, the refractive index of the extraordinary component of light, having polarization direction parallel to the optic axis direction, is a function of the angle between the propagation vector of the beam with the optic axis. This phenomenon in addition to the walk-off between the Poynting vector direction and the wave propagation vector direction results in a spatial shift between the extraordinary ray (e-ray) and ordinary ray (o-ray), when the direction of propagation is not parallel or normal to the optic axis direction [42, 48]. This is called unfolding due to birefringence.

A birefringent crystal for which the optic axis is normal to the refracting surface, the angle made by the ordinary ray ($\theta_o$) and extraordinary ray ($\theta_e$) inside the crystal is given by [48],

$$\tan\theta_o = \frac{\sin\theta}{\sqrt{n_o^2 - \sin^2\theta}} \quad \text{and} \quad \tan\theta_e = \frac{n_o \sin\theta}{n_e^2\sqrt{n_e^2 - \sin^2\theta}}, \quad (3)$$

where the crystal is rotated at an angle $\theta$ about the incident beam from the orientation of normal incidence on the refracting surface. $n_o$ and $n_e$ are the ordinary and extraordinary refractive indices respectively.

Using Eq. (3), the spatial separation between the o-ray and e-ray at the crystal output plane can be written as, $\Lambda = t(\tan\theta_o - \tan\theta_e)$, where '$t$' is the thickness of the crystal. Also because of different path taken by the e-ray and o-ray and due to their different refractive indices inside the crystal, there is an optical path length difference between them. The optical path length difference at a normal plane to the propagation direction after the crystal is given as, $\delta l = t(n_o \sec\theta_o - n_e \sec\theta_e) - \Lambda \sin\theta$ and the corresponding phase difference is given as $\Delta = (2\pi/\lambda) \times \delta l$.

The calculated values of the transverse spatial separation ($\Lambda$) and the dynamic relative phases ($\Delta$) between the e-ray and o-ray of an unfolded OV beam inside a c-cut birefringent crystal, for increasing crystal rotation from normal incidence, is shown in Fig. 1 (a) and (b) respectively. In (b) a part of the plot (shown in red dashed rectangle) is magnified and replotted for visual clarity. Here the $n_o$ and $n_e$ of the YVO$_4$ crystal is taken (1.993 and 2.215 respectively for 633 nm wavelength) and the length of the crystal is taken as 15 mm (i.e. $t$ = 15 mm). It is clearly observed that, $\Lambda$ increases linearly with increasing crystal angle ($\theta$). Also, for very small change of $\theta$ after about 2 degrees the variation of $\Delta$ with $\theta$ is almost linear within one cycle of $2\pi$ phase change.

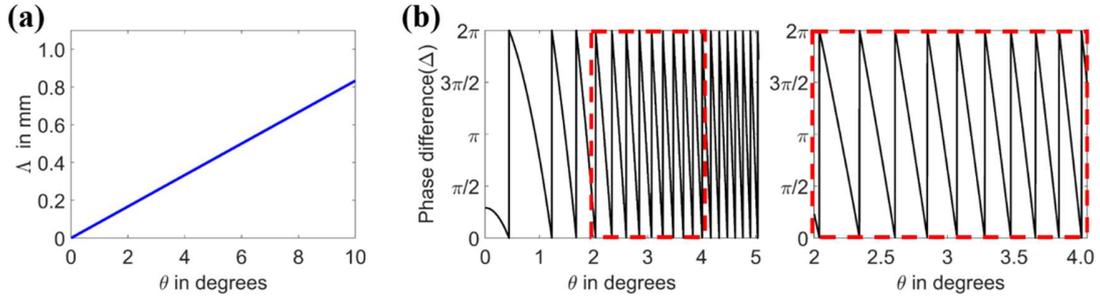

**FIG. 1.** **(a)** Transverse spatial shift ($\Lambda$) and **(b)** dynamic phase difference ($\Delta$) between the e-ray and o-ray due to propagation inside a c-cut birefringent crystal with optic axis making an angle $\theta$ with the propagation direction. Part of the plot in (b) shown by red dashed rectangle is replotted after magnification for visual clarity.

### C. Stokes fields due to unfolding of the FOV beams:

The state of polarization at a point across a beam is characterized by four Stokes parameters ($s_0$, $s_1$, $s_2$, $s_3$), which can be obtained from the six intensity images I ($\beta$, $\gamma$) where $\beta$ and $\gamma$ are the angle of the quarter-wave-plate and polarizer respectively [42-44]. From the normalized Stokes parameters ($S_0$, $S_1$, $S_2$, $S_3$), three complex scalar Stokes field ($S_{12}$, $S_{23}$, $S_{31}$) can be constructed by $S_{mn} = S_m + iS_n$. The orientation field i.e. orientation of the major axis ($\psi$) of the polarization ellipse is obtained from argument of the $S_{12}$ and the relative phase ($\delta$) between the eigen-polarized components is obtained from argument of the $S_{23}$ as follows [43].

$$\psi = \tfrac{1}{2}[\tan^{-1}(\tfrac{S_2}{S_1})]$$
$$\varphi = \tfrac{1}{2}\tan^{-1}\left(\frac{(S_1^2 + S_2^2)^{1/2}}{S_3}\right) \quad (4)$$
$$\chi = \frac{\pi}{4} - \varphi$$
$$\delta = \tfrac{1}{2}[\tan^{-1}(\tfrac{S_3}{S_2})]$$

The angular parameters ($\psi$, $\varphi$) span the parameter space of Poincaré sphere (PS) which has radius $S_0 = 1$ and $S_1$, $S_2$, $S_3$ as three axes. $S_3 = \pm 1$ represent the poles and $S_1S_2$ plane is the equatorial plane. The parameter $\chi$ expresses the ellipticity of the polarization ellipse. The polarization ellipse and the corresponding SoP parameters on the PS is shown in Figs. 2. (a) and (b), respectively. C-point singularity is denoted by $\chi = \pm \pi/4$ i.e. $2\varphi = (0, \pi)$ where $\psi$ is undefined and projects to the pole of PS. The L-line is expressed by $S_3 = 0$ and given by $2\varphi = \pi/2$. The upper hemisphere represents left handed ellipses and the lower hemisphere represents right handed ellipses separated by the handedness singularity L-line along the equator.

### D. Geometric phase associated with transverse polarization structure:

The SoP along any closed path on the physical space across the inhomogeneously polarized vector vortex beam also traces a closed loop on the spherical parametric surface of the PS. When the circular path is taken around the C-point singularities in the physical space, the PS contour always encloses the $S_3$ axis. The half of the oriented area (solid angle) enclosed by the PS contour gives the value of the associated GP [37]. For a vectorial electric field distribution represented by $\mathbf{E}(r)=E(r)\,\mathbf{e}(r)$, the geometric phase is given by,

$$\Omega_G = \text{Im}\oiint [\mathbf{e}^*.(\nabla)\mathbf{e}].\,dr = \oiint (1 \mp \cos(2\varphi))\,d\psi \qquad (5)$$

where $\mathbf{e}(r)$ is the unit vector representing the polarization state at point '$r$' on the orthogonal plane about the beam propagation. The expression of $\Omega_G$ in terms the SoP parameters is employed to calculate the area on the PS surface within the PS contour. The process is indicated in Fig. 2(c) and explained as following.

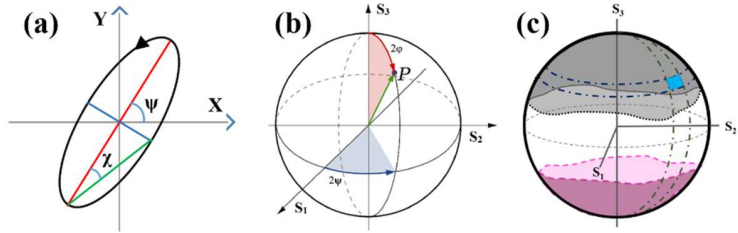

FIG. 2. (a) Polarization ellipse and (b) Poincaré sphere on which the SoP is represented by the point P with parameters that correspond to its semi-polar and semi-azimuthal angles on the sphere. (c) The symbolic representation of the trace of the SoP along a circle around star and lemon C-points (shown in black and magenta dotted line respectively) on the Poincaré sphere and to calculate the oriented surface area (gray and magenta) enclosed by the projected contours and the poles.

From the SoP parameter $\psi$ and $\varphi$ calculated along the circular path around the C-point, the angular coordinate of the points on PS contour is known. As the north hemisphere of the PS represents left-handed polarization and the south hemisphere represents right-handed polarization, the area enclosed by the PS contours is given by the area within the north pole and the contour traced by black dotted line and the area within south pole and the contour traced by magenta dashed lines. In calculating the GP enclosed by the PS contours, the geometric angular coordinates ($2\varphi$, $2\psi$) of the data points on PS is employed in numerically integrating or summing the infinitesimal spherical surface elements within the contours following Eq. (5). In Eq. (5), the '-ve' sign is to be taken for the PS contour around the north pole (in this study, around the star C-point) and '+ve' sign is for the PS contour around the south pole (here it is around lemon C-point).

### III. EXPERIMENTAL DETAILS:

The blazed CGH with different fractional order MSED is displayed on a reconfigurable and dynamically controllable spatial light modulator (SLM). The SLM employed is a phase only reflective SLM with pixel pitch of 8μm (Pluto from Holoeye). The diffracted beam from the fractional order CGH results in the envisaged scalar FOV beam in the first diffraction order [25]. Conversion of phase singularities to polarization singularities is achieved by unfolding the generated FOV beams inside a birefringent crystal in two eigen-polarization component as shown in Fig. 3.

First, the generated scalar FOV beam is collimated and polarization is made diagonal. Then the beam is made to pass through a c-cut uniaxial birefringent Yttrium vanadate ($YVO_4$) crystal (from Altechna) placed on a micro-radian precision rotational stage. The dimension of the crystal is 15 mm(length) X 10 mm(width) X 10 mm(height). The optic axis of the crystal which is normal to the refracting surface, lies in the horizontal y-z plane but is rotated about x axis and making an angle θ with z axis. The FOV beam is unfolded upon entering the crystal surface and are spatially separated at the output plane in two orthogonally polarized eigen-components. The spatial separation of the beam can be controlled by the angle (θ) of optic axis about the z axis, as shown in Fig. 1(a). Any change of this crystal angle will also introduce a longitudinal optical path difference between the two orthogonal polarized components at the crystal output plane. The corresponding relative dynamic phase difference is found and demonstrated to be most dominant factor among all the factors controlling the nature of the generated vector vortex beams.

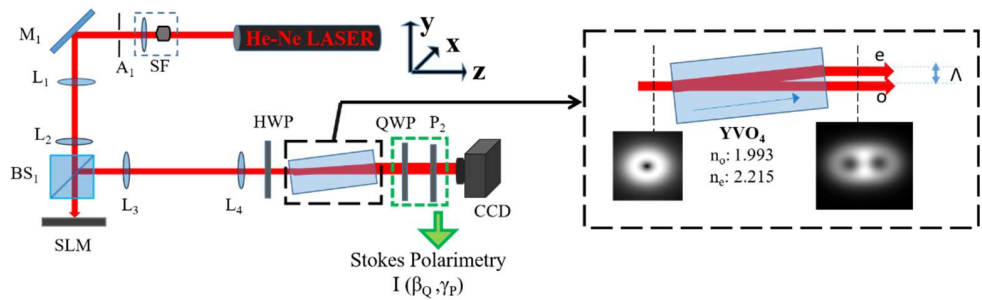

**FIG. 3.** Generation of spatially inhomogeneous polarized vector vortex beam via unfolding of a fractional OV beam in a c-cut birefringent ($YVO_4$) crystal. SF: spatial filter, SLM: spatial light modulator, HWP/QWP: half/quarter-wave-plate, P: polariser.

To measure and vary the longitudinal relative phase (Δ) between the two eigen-polarization components, first the orientations of the crystal giving rise to effective zero (modulo π) phase difference are found by rotating the crystal within a crossed polarizer (kept in diagonal (D) - anti-diagonal (A) configuration). This is not shown in figure. At these orientations of the crystal, the output intensity profiles show a dark line at the centre of the beam where the phase difference between the eigen-polarization components is zero modulo π. This concept is taken from the weak measurement of very small amount of birefringence induced spatial walk-off of a polarized beam [46]. Assuming the phase difference to be approximately linear with θ for very small change between two consecutive effective zero phase difference configurations, any arbitrary relative phase difference (Δ) between the eigen-polarization components can be employed within the experimental accuracy and in compliance to the precision of the rotational stage. The beam size is measured using two-dimensional Gaussian fitting and the beam separation is calculated using both the centre of the fitted Gaussian and centre of intensity calculation at the CCD plane. The beam radius (w) of the Gaussian beam is found to be $1.0 \pm 0.04$ mm.

The output state of polarization (SoP) at a particular orientation of the crystal is measured by spatially resolved Stokes polarimetry using a quarter-wave-plate (QWP) and polarizer ($P_2$) combination. The spatial distribution of SoP is measured with varying beam separation ($\Lambda$) and dynamic relative phase ($\Delta$) for different fractional order ($\alpha$) of the OV beams. The measured relative phase ($\delta$) between the eigen-polarizations using Eq. (4) will be different from the given longitudinal relative phase ($\Delta$) for any beam that does not have a planar wavefront (i.e. for $\alpha \neq 0.0$). So, to measure the value of $\Delta$, the Gaussian beam is used and $\delta$ is calculated. (For planar wavefront Gaussian beam $\delta = \Delta$).

## IV. RESULTS AND DISCUSSIONS:

The unfolding of the FOV beams inside the birefringent crystal in to two orthogonal polarization components, results in the output SoP as shown in Fig. 4. In the SoP distributions, red and green ellipses imply right and left-handed polarization states respectively and the blue line represents the L-line. The orientation fields ($\psi$) with polarization streamlines and corresponding spatial distribution of relative phase ($\delta$) derived from the constructed Stokes field using Eq. (4) are also shown.

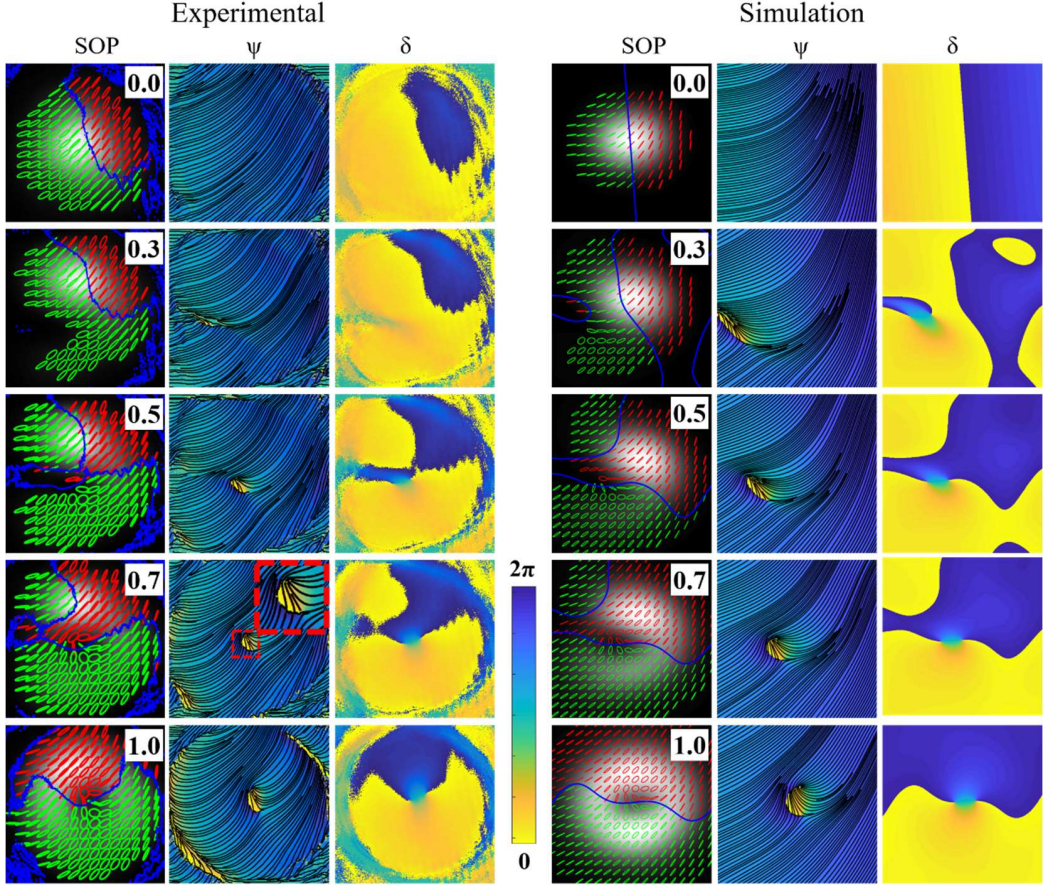

**FIG. 4.** Unfolding of FOV: Experimental and simulation plots for beam separation $\Lambda=0.4w$ and zero longitudinal phase difference ($\Delta=0$). SoP across the beam where red and green ellipse represents right and left handedness and the blue line is L-line. Orientation of the ellipse major axis ($\psi$) with the singularities representing C-points and polarization streamlines revealing the index of the C-points (In the experimental plot for $\alpha = 0.7$, the inset shows the region inside the dashed square around the C-points, magnified for visual clarification of lemon and star C-points). Transverse

distribution of relative phase (δ) between the two eigen-polarization components as a result of unfolding.

The C-points and L-lines are also obtained from the measured Stokes parameters. The points of intersection of the zero contours of $S_1$, $S_2$ as obtained from the singularities of the orientation field $\psi$ = Arg ($S_{12}$) gives the C-points. The L-lines are drawn as zero contours of $S_3$. The singularities of the orientation field clearly show the C-points. The nature of the C-point singularity (lemon or star) is revealed by the polarization streamlines. Three streamlines merging at the C-point implies a star singularity and a single streamline ending at the C-point implies a lemon singularity. The other boundary type monstar singularity is known to be rare in general [8], and is not observed in this study.

The SoP distribution at the output plane is also simulated by superposing two orthogonal polarized FOV beams using the calculated and measured beam separation and dynamic relative phase from experiments and using Eq. (1) and Eq. (2). It is observed that the number and type of C-point singularities depends on the fractional order of spiral phase dislocation the input beam is imprinted with, the spatial separation of the unfolded eigen-polarized components at the output plane and also on the longitudinal relative phase difference between the components. In the presented results in Fig. 4, the beam separation (Λ) is 0.4w (where w is the Gaussian beam radius) and the longitudinal relative phase is zero. The experimental results are well supported by simulation.

Fig. 5 shows the variation of the three angular parameters ($\psi$, $\varphi$ and $\chi$) characterizing the SoP along a circle (radius 100μm) centred on the C-points with the azimuthal angle of the local coordinate frame where the origin is placed on the C-point. The variation of the parameter $\psi$ when going counter-clockwise along the circle centred on a C-points, dictates the sign of that C-point index. The sign of the C-point index is positive for increasing $\psi$ representing lemon and negative for decreasing $\psi$ representing star C-points. This experimentally measured and calculated values of the SoP parameters are also used later to plot the contours on the PS and calculating the oriented area (solid angle) enclosed by the PS contours.

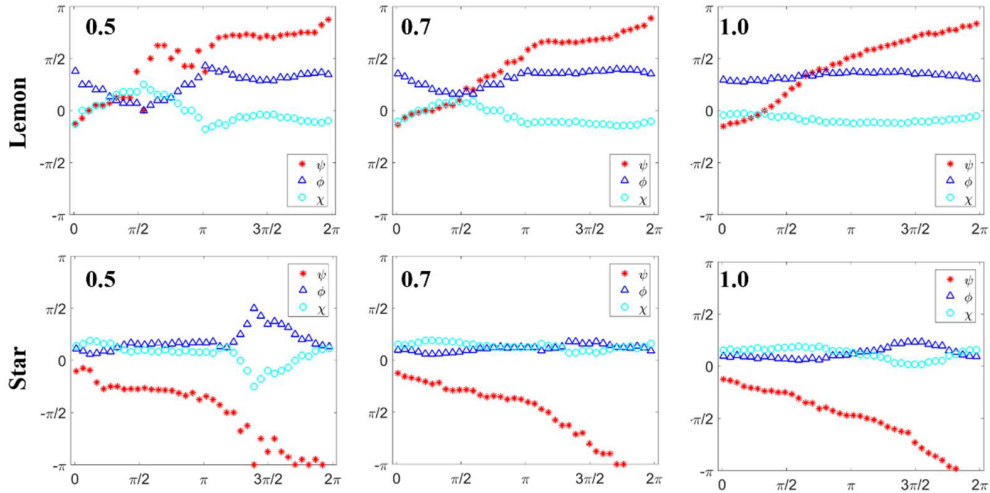

**FIG. 5.** Variation of the SoP parameters ($\psi$, $\chi$, $\varphi$) along the circle around the two C-points for the vector vortex beams (shown in Fig. 4) generated by unfolding of the FOV beams with α = 0.5, 0.7 and 1.0

It is observed that for any $0.5 \leq \alpha \leq 1.0$, when the dynamic relative phase (Δ) due to longitudinal path difference is zero (modulo π) between the two eigen-polarized beams, the number of C-points within the beam cross-section is always two: one right-handed lemon and another left handed star C-point singularities. Thus the total index which is the sum of the product of the sign of the C-point index ('+' for lemon and '−' for star) and sign of the handedness index ('+' for right-handed and '−' for left-handed) [34], matches with the total topological charge of the scalar vortex as both equals positive two. The C-points enter within the beam cross-section through the opening of L-line formed near the lower intensity regions corresponding to the edge dislocation of the grating hologram at α = 0.5. With increasing α in fractional steps, the C-points move towards the beam centre and reaches the centre at α = 1.0.

When the relative phase difference (Δ) between the eigen-components is changed by a very small rotation of the crystal, the SoP across the beam are observed to pass through a rapid cycle of change. In the following Fig. 6, the effect of beam separation (Λ) and relative phase (Δ) on the SoP is shown in (a) and (b) respectively for FOV of α = 0.5. In (a) for demonstrating the effect of change in Λ, Δ is fixed at $2\pi/3$ and in (b) for showing the effect of Δ, Λ is fixed at 0.4w. It is clearly observed that the nature of the Poincaré beams greatly depends on the value of Δ.

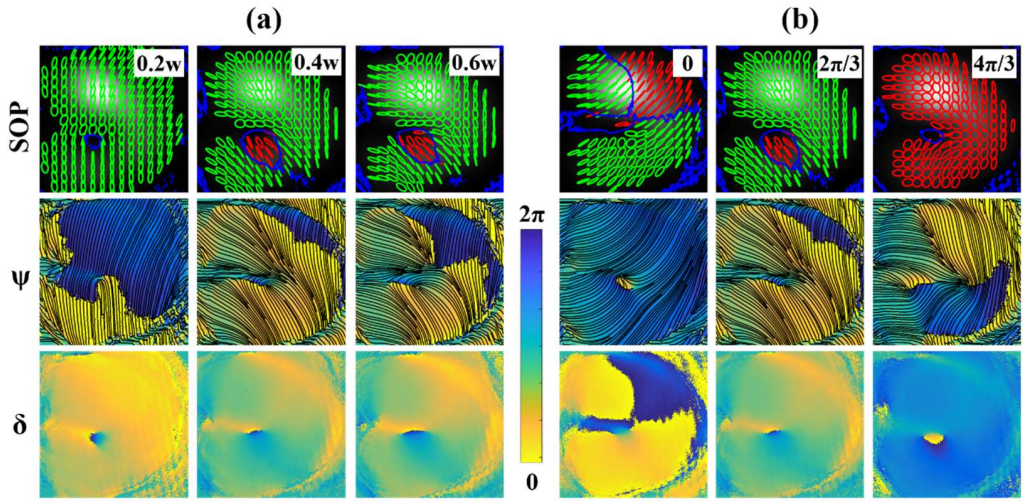

**FIG. 6. (a)** Effect of beam separation (Λ) for a fixed relative phase (Δ=$2\pi/3$) and **(b)** effect of relative phase (Δ) for a fixed beam separation (Λ=0.4w) on the SoP distribution of generated Poincaré beams via unfolding of an OV beam (α=0.5).

Below α = 0.5, there are no C-points across the beam cross-section. So if all the SoP across the beam is projected on the PS, it will not be sufficient to span the whole surface of the PS as this SoP distribution do not include the poles. In this cases, the generated Poincaré beam is a partial Poincaré beam. For $0.5 \leq \alpha \leq 1.0$, at Δ=0, there are two C-points that are separated by a L-line. Though the SoP across the beam now span the poles of the PS, the open L-line indicates that it is still insufficient to cover the whole surface of the PS. With increasing relative phase (Δ) at a fixed beam separation, the L-line turns into a closed loop enclosing one C-point after some critical value of Δ. From simulation, this critical value is found to be of the order of $\pm \pi/18$. It is difficult to get this critical value from experiment as it requires very small and precise control of the relative phase, but the close matching of the experimental and simulation results indicates that the same can be obtained if a motorized precise rotational stage to be employed in the experiment. When the L-line encloses one C-point forming a closed loop within the beam cross-section, then the SoP across the beam span the full PS and the beams now become a full Poincaré beam. So with varying α and Λ, it is possible to transform a partial Poincaré beam in to a full Poincaré beam via unfolding.

It is also observed that a new pair of C-points: one star and one lemon, of the same handedness may arise in the beam cross section after some critical beam separation. The new pair of C-points both being of same handedness does not contribute to the total index as they have opposite sign of the C-point index. And the L-line encloses one out of two (or four) C-points: either one right-handed lemon or a left-handed star. The following table (Table 1) summarizes the number and index of the C-points observed.

The L-line is an open line intersecting the beam in half when the relative phase is around 0 or $\pi$ and move very rapidly across the beam. This correspond to the degenerate case for the L-line singularity as a relative phase of 0 or $\pi$ between two orthogonal linear polarization components gives a linear polarization. For any other relative phase, the L-line very rapidly takes a closed loop form and encloses a C-point singularity near the beam centre as explained earlier. When the phase difference approaches near $\pi/2$ or $3\pi/2$, the SoP across most of the beam cross-section is close to circular polarization and the beam is found to be embedded with three C-points. In this cases, the new C-points move very rapidly across the beam with slight change of $\Delta$. This is denoted by a 'degenerate' {d} condition in table 1, as a relative phase of $\pi/2$ between two orthogonal polarization gives circular polarization states.

TABLE 1: Number and index of C-points for some relative phase ($\Delta$) and spatial separation ($\Lambda/w$) of the eigen-polarization components: (L, S represents lemon and Star C-points respectively; Ł and $ represents new C-points having opposite signed index that do not contribute in the total topological index)

| $\Delta \rightarrow$ $\Lambda/w \downarrow$ | 0 | $\pi/2$ | $2\pi/3$ | $3\pi/2$ | $4\pi/3$ |
|---|---|---|---|---|---|
| 0.2 | L, S | d | L, S | d | L, S |
| 0.4 | L, S | d | L, S | d | L, S, Ł, $ |
| 0.6 | L, S | d | L, S, Ł, $ | d | L, S, Ł, $ |

To study the nature of the Poincaré beams, the SoP around the C-points are projected on the PS. As already mentioned, on the PS the poles are C-points and the equator spans the L-line and separates the right-handed and left-handed SoP. So, SoP along any circle around one of C-points traces a circle around one of the pole. The position, orientation and tilt of the contour on the PS, about the equator is a qualitative measure of the local structure of the C-point singularity as it describes the variation of the SoP around the C-point [37, 49]. The unfolding area, defined by the region at the boundary of which the SoP projects to the half of diametrically opposite SoP on the PS from the SoP at the centre of the beam [50]. This is a direct measurable quantity from the relative phase ($\delta$). The SoP and the projected PS contours from the SoP along circles of radii 100 μm centred on the C-points are shown in Figs. 7 and 8 for varying $\alpha$, $\Lambda$ and $\Delta$ and explained below.

In Fig. 7, the effect of varying the beam separation ($\Lambda$) for a particular relative phase ($\Delta = 0$) is shown. The lemon and star C-points are shown with a magenta circle and black star whereas the circle along which the SoP parameters are calculated, are shown in magenta and black line. The projection of SoP around the lemon and star C-points on the PS is shown in magenta circles and black stars respectively. With increasing separation, the inhomogeneity of the SoP around the C-points decreases and the circle on the PS transits towards a loop parallel to the equator of the sphere. The tilt of the projected circle on the Poincaré sphere with respect to the equator is a measure of the anisotropy of the SoP around the C-points [49]. In Fig. 8 the effect of varying relative phase ($\Delta$) for a particular the beam separation ($\Lambda = 0.4w$) is shown.

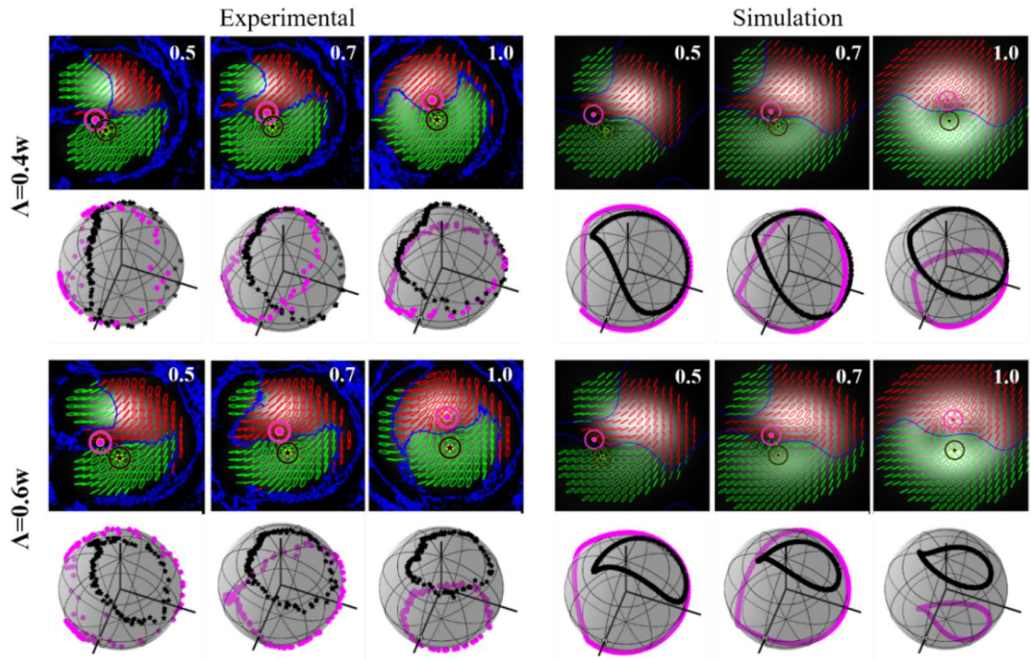

**FIG. 7.** Projection of SoP along a circle (radius 100 μm) around the C-points on the Poincaré sphere for beam separation: Λ=0.4w and 0.6w (w is the Gaussian beam radius) for α= (0.5, 0.7 and 1.0); Δ = 0.

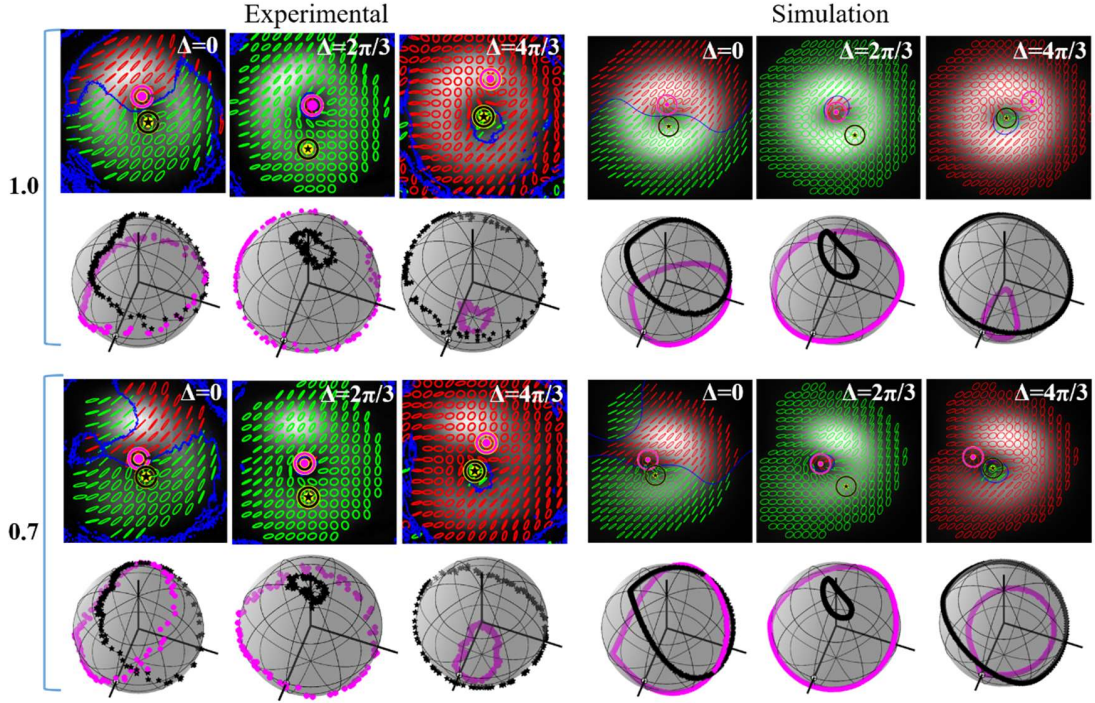

**FIG. 8.** Projection of SoP along a circle around the C-points on the Poincaré sphere for varying relative phases: Δ=0, 2π/3 and 4π/3; For α= 0.7, 1.0; Λ=0.4w (w is the Gaussian beam radius).

The enclosed spherical surface area by the Poincaré sphere (PS) contour is a measure of the GP associated with the C-point singularity [37]. The variation of the GP due to the PS contours calculated using Eq. (5) for different fractional values of α are shown with varying relative phases in Fig. 9 (a). Though the actual amount of GP is dependent on the radius of the circular path chosen around the C-points but the qualitative nature of the variation with α is same. The variation of GP for taking different radii of the circle around the C-point is also shown in Fig. 9(b).

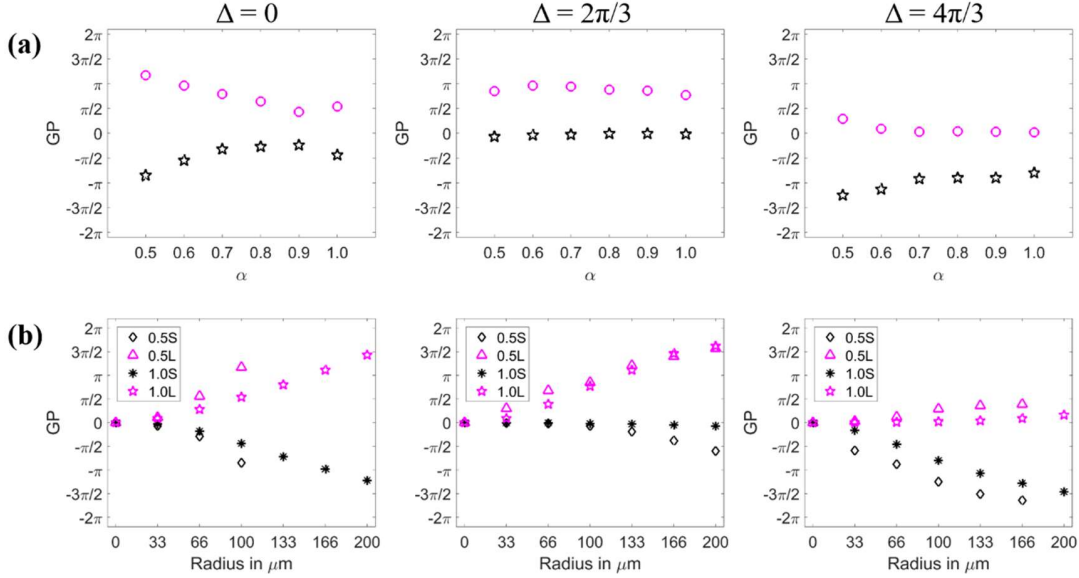

**FIG. 9. (a)** Variation of the GP with varying α for Δ = 0, 2π/3 and 4π/3 at Λ = 0.4w. Radius of the circle is 100 μm.
**(b)** Variation of GP with different radii of circles around the C-points for Poincaré beams from α = 0.5 and 1.0; S and L in legends implies star and lemon C-points respectively.

In Fig. 9(a), magenta circle represents GP for lemon C-points whereas black star represents GP for star C-points. The plot of GP variation with α for different value of Δ reveals that variation is maximum for Δ = 0 and least for Δ = 2π/3 where GP is almost independent of α. From Fig. 9(b), it is observed that, variation of GP with radius of the circle around C-points, are same for the lemon and star C-points at Δ = 0. But the variation is more rapid for lemon C-points at Δ = 2π/3 whereas it is more rapid for star C-points at Δ = 4π/3. This is because the L-line encloses the lemon C-point for Δ = 2π/3 and encloses the star C-point for Δ = 4π/3.

For α = 0.5 and Δ = 0, the C-points are near the low intensity regions due to the phase dislocation line as can be seen in Fig. 7. So in this case, the circle with increasing radii around the C-points cannot be increased beyond some value when the circle goes beyond the beam cross-section. In Fig. 9. (b), for Δ = 0 the radius of the circles around the C-points for α = 0.5 are thus increased to maximum of 100 μm, beyond which the SoP parameters along the circle could not calculated.

## V.  CONCLUSIONS:

To summarise the investigation, scalar fractional optical vortex beams are generated using computer generated holograms embedded with fractional helicoidal phase steps. By unfolding the generated diagonally polarized FOV beams inside a uniaxial c-cut birefringent $YVO_4$ crystal, Poincaré beams are generated at the output of the crystal due to the spatial walk-off of the eigen-polarization components. The output transverse polarization distribution reveals that the scalar phase singularities of the input beam are transformed to polarization singularities. The complex phase structure of the FOV beams, the transverse spatial separation of the eigen-polarization components and their longitudinal relative phases, are major factors in determining the complex polarization structure of the generated Poincaré beams. The states of polarization around the C-points are projected on the Poincaré sphere. The fractional order of the of the CGH determines the phase structure and local anisotropy of the vortices which themselves can be mapped to a morphology sphere in analogy to the Poincaré sphere [45].

In addition to the fractional order of the CGH, the spatial separation and the relative phase of the two eigen-polarization components determine the number and index of C-points within the beam cross-section, the partial or full spanning of the Poincaré sphere by the projected states of polarization and the area of the unfolding region of the unfolded FOV beams. The geometric phase due to the Poincaré sphere contour is calculated and the variation of geometric phase with all the different determining parameters are investigated. This study can be useful in beam shaping, engineering of polarization and point spread functions for imaging, in metrology and laser materials processing etc. The robust nature of the geometric phase enhances the suitability of the generated Poincaré beams in quantum applications. The classical 'entanglement' of the polarization and spatial-mode degrees of freedom in Poincaré beams is also another promising feature to be useful in classical realization of some algorithms which only require the entanglement feature of quantum systems. This study may be further helpful in realizing the controlled degree of classical entanglement which requires further investigation of these beams.

**ACKNOWLEDGMENTS**

This research work is supported by Department of Science and Technology (DST), India with grant number (INSPIRE Faculty Award/2013/PH-62) and Indian Institute of Technology Kharagpur (IITKGP), India with grant number (IIT/SRIC/PHY/VBC/2014-15/43).

**ORCID iDs**

Satyajit Maji: https://orcid.org/0000-0002-9432-7615
Maruthi M. Brundavanam: https://orcid.org/ 0000-0003-0442-3599